\pdfoutput=1
\documentclass[prd,preprint,floats,floatfix,aps,nofootinbib,preprintnumbers]{revtex4}
\usepackage{graphics,graphicx,psfrag}
\usepackage{amsmath,amssymb}
\usepackage[sort&compress]{natbib}
\usepackage{bm}	
\usepackage{verbatim}
\usepackage{multirow}
\usepackage{subfigure}
\usepackage{ulem}
\usepackage{color}
\usepackage[linktocpage=true,bookmarksnumbered=true,colorlinks=true,plainpages,linkcolor=blue,citecolor=red]{hyperref}
\usepackage[utf8]{inputenc}  
\usepackage{soul}	
\soulregister\cite7
\soulregister\ref7
\soulregister\pageref7
\def\beq{\begin{equation}}
\def\eeq{\end{equation}}
\def\bea{\begin{eqnarray}}
\def\eea{\end{eqnarray}}
\def\bealn{\begin{eqnarray}}
\def\eealn{\end{eqnarray}}
\def\nn{\nonumber}
\newcommand{\pbrac}[1]{\left( #1 \right)}
\newcommand{\tbrac}[1]{\left[ #1 \right]}

\def\lag{\mathcal{L}}
\def\psa{\mathcal{A}}
\def\vevh{v_{h}}
\def\vevs{v_{s}}

\newcommand{\fb}{\text{fb}}

\begin{document}
\preprint{
{\vbox {
\hbox{\bf MSUHEP-160407}
\hbox{\today}
}}}
\vspace*{2cm}

\title{Diphoton Resonances in the Renormalizable Coloron Model}
\vspace*{0.25in}   
\author{R. Sekhar Chivukula$^1$}
\email{sekhar@msu.edu}
\author{Arsham Farzinnia}
\email{farzinni@msu.edu}
\author{Kirtimaan Mohan$^1$}
\email{kamohan@pa.msu.edu}
\author{Elizabeth H. Simmons$^1$}
\email{esimmons@msu.edu}
\affiliation{\vspace*{0.1in}
$^1$ Department of Physics and Astronomy\\
Michigan State University, East Lansing U.S.A.
}
\vspace*{0.25 in} 


\begin{abstract}
\vspace{0.5cm}
\noindent

The renormalizable coloron model, which has previously been shown in the literature to be consistent with a wide array of theoretical and precision electroweak constraints, includes a pair of spinless bosons (one scalar, one pseudoscalar).  We show that either of them, or both together if they are degenerate, could be responsible for the diphoton resonance signal for which both CMS and ATLAS have seen evidence.  Because either of these bosons would be produced and decay through loops of spectator fermions, the absence of signals in dijet, $t\bar{t}$, and electroweak boson pair channels is not a surprise.

\end{abstract}

\maketitle

\section{Introduction}
\label{sec:intro}

The LHC has seen an indication of a diphoton resonance at 750 GeV in the CMS \cite{CMS:2015dxe,CMS-PAS-EXO-16-018,Moriond-CMS} and ATLAS \cite{ATLAS-diphoton,Moriond-ATLAS,ATLAS-CONF-2016-018} experiments.  Many potential classes of new physics explanations have been catalogued in refs. \cite{Franceschini:2015kwy,Staub:2016dxq} and a large number of papers have suggested additional possibilities.  It has turned out to be challenging to create models that are consistent with the properties of the resonance, do not violate constraints established by previous experiments, and do not include unreasonably large numbers of new particles.  This letter proposes a new explanation for the resonance that has the virtue of being part of a model that is already established to be consistent with other existing constraints on new electroweak physics.

A challenge in proposing new states to explain the diphoton signal is they must be readily enough produced to agree with the observed cross-section while evading constraints imposed by the lack of observed signals in dijet, $WW$, $ZZ$ channels at 750 GeV. Following the prescription used in \cite{1512.04939}, for a narrow resonance ${\mathcal R}$, the resonance production cross section times diphoton branching ratio needed to explain the signal at the 13 TeV LHC is estimated to be\footnote{ A more recent estimate, including initial data from the 13 TeV run, yields a somewhat lower value for the estimated cross section, with a central value of 4.8-5.5 fb$^{-1}$ \cite{Franceschini:2016gxv}. These revised cross section estimates are within the range considered here.}
\begin{equation}
\sigma(pp \to {\mathcal R} \to \gamma \gamma )= 6.26 \pm 3.32\  \fb\ .
\label{eq:signal}
\end{equation}
{ In what follows, we will consider regions of parameter space that produce 13 TeV signal cross sections of between 3 and 9.4 fb$^{-1}$.}
At the same time, exclusions on a $750$ GeV resonance ${\mathcal R}$ decaying to other standard model (SM) particles are determined by using the following set of values taken from {8 TeV LHC} experimental analyses:
\begin{description}
\item \qquad\qquad $\sigma({pp \to {\mathcal R} \to Z \gamma }) < 8.2\ \fb$  \cite{Aad:2014fha}, \qquad $\sigma({pp \to {\mathcal R} \to W^{+} W^{-} }) < 37\ \fb$ \cite{Aad:2015agg},
\item \qquad\qquad $\sigma({pp \to {\mathcal R} \to Z Z }) < 19\ \fb$ \cite{Aad:2015kna}, \qquad $\sigma({pp \to {\mathcal R} \to g g }) < 2200\ \fb$ \cite{CMS-PAS-EXO-14-005},
\item \qquad\qquad $\sigma({pp \to {\mathcal R} \to t \bar{t}  }) < 700\ \fb$ \cite{Aad:2015fna}.
\end{description}

We will show in this note that the observed diphoton resonance could be due to scalar and pseudoscalar states in the renormalizable coloron model \cite{Bai:2010dj}, a model that has been previously studied in the literature \cite{Chivukula:1996yr,Hill:1993hs,Dicus:1994sw} and is already known to be consistent with electroweak precision constraints and theoretical constraints \cite{Chivukula:2013xka,Chivukula:2014rka,Chivukula:2015kua}.  More specifically, either the scalar or pseudoscalar state in the model could be responsible for the diphoton signal -- or the two states could be degenerate and jointly responsible.  

The model contains an extended color gauge group and the new scalar and pseudoscalar arise as part of the sector that spontaneously breaks the extended group down to standard QCD.  In consequence, the new scalars do not couple directly to quarks and their mixing with the Higgs (which could induce a small indirect coupling to quarks) must be nearly zero to comport with precision electroweak data.  Rather, the new scalars couple to spectator quarks that help cancel gauge anomalies in the theory. Gluon pairs coupled to loops of these spectators allow for s-channel production of the scalars at LHC; photon pairs likewise coupled to spectator loops allow for decay.  Because production and decay are all occurring through loop-level processes, the dijet, $WW$,  $ZZ$, and $Z\gamma$ rates can be small enough to be consistent with the LHC constraints. 

One last key element arises because the extended color sector yields an octet of massive coloron bosons that would be visible at LHC.  The most recent limits on colorons have been set by CMS, which finds that the coloron mass must exceed 5.1 TeV \cite{Khachatryan:2015dcf,Chivukula:2011ng,Chivukula:2013xla}. Because the scalars are part of the color symmetry-breaking sector, their vacuum expectation value ($\vevs$) is linked to the mass of the coloron~\cite{Chivukula:2013xka}; hence, the new limit on the coloron mass means that $\vevs$ must be at least $1.7$ TeV.\footnote{As explained below, in what folows we will use $\vevs=2$ TeV for illustration. Larger values are also allowed, though the fermion content of the theory must be adjusted accordingly to accomodate the observed diphoton signal.}

Putting all of this information together, we find that the renormalizable coloron model is consistent with all of the data if one adds a few weak-singlet spectators to complement the weak-doublet spectators in the original model. The presence of the additional spectators enables the new scalar and/or pseudoscalar to be visible as a diphoton resonance without producing dijet, $WW$, or $ZZ$ events that would contravene the LHC bounds.  Moreover, the addition of weak-singlet scalars leaves the model still in agreement with precision electroweak constraints and has only a small impact on the details of how the other theoretical constraints ({\it e.g.} triviality) are satisfied.\footnote{A previous paper in the literature~\cite{Liu:2015yec} suggested coloron decay to diphoton + jet might be the source of the LHC diphoton signal.  That work did not include any contribution from scalar or pseudoscalar states which are the focus of the present work. Moreover it assumed a coloron mass of $2$ TeV, which is now well below the LHC’s exclusion limit of $5.1$ TeV.} 

In the rest of this letter, we lay out the details of how the diphoton resonance appears in the renormalizable coloron model, what model components are necessary to ensure compliance with all phenomenological constraints, and what open questions should be studied if the resonsance is confirmed by additional LHC data.  In section 2, we briefly review the elements of the renormalizable coloron model.  Section 3 presents our calculations related to the diphoton signal observed at LHC.  Section 4 presents a discussion and summarizes our conclusions.

\section{Elements of the Model}
\label{sec:model}

\subsection{Bosonic Sector}
\label{subsec:boson}

The renormalizable coloron model is based on an extended $SU(3)_{1c} \times SU(3)_{2c}$ gauge symmetry, where color $SU(3)_C$ is identified with the diagonal subgroup of the larger group.
The extended group is broken down via the expectation value of a $(3,\bar{3})$ scalar,
$\Phi$, which may be decomposed into gauge eigenstates of QCD as follows
\begin{equation}\label{Phi}
\Phi = \frac{1}{\sqrt{6}} \pbrac{v_{s} + s_{0} + i {\mathcal A}} {\cal I}_{3\times 3} + \pbrac{G^a_H + i G^a_G}t^a \qquad \pbrac{t^a \equiv \lambda^a/2} \ .
\end{equation}
Here the $t^a$ are the generators of $SU(3)$, $\vevs$ is the magnitude of the vacuum expectation value breaking the extended color symmetry, $s_0$ and ${\mathcal A}$ are singlet scalar and pseudoscalar fields, and $G^a_H$ and $G^a_G$ are color-octet scalar and pseudoscalar fields. The $G^a_G$ fields are absorbed by the massive color octet vector fields, the colorons, after symmetry breaking; the  $G^a_H$ remain as physical states of the theory and their phenomenology has been studied in \cite{Bai:2010dj,Hill:1993hs}. The $s_0$ (after mixing with the Higgs field, as described below) and the ${\mathcal A}$ fields are candidate states for a diphoton resonance at 750 GeV. 

The model also includes a color-singlet weak-doublet Higgs field ($\phi$), whose neutral component develops a vacuum expectation value $v_h/\sqrt{2}$ (with $v_h \approx 246$ GeV) and is responsible for electroweak symmetry breaking. The scalar component of the Higgs field that remains in the spectrum after electroweak symmetry breaking ($h_0$) mixes with the $s_0$ scalar via a mixing angle $\chi$ to form mass eigenstate scalars
\begin{align}
s & = \sin\chi \, h_0 + \cos \chi\, s_0~,\\
h & = \cos \chi\, h_0 - \sin\chi\, s_0~.
\end{align}
An analysis of the model's full scalar potential phenomenology is given in \cite{Chivukula:2013xka,Chivukula:2014rka,Chivukula:2015kua}; one key result is that the value of $\sin\chi$ is constrained to be very small ($\lesssim 0.1$).

The coloron mass in this model is given by
\begin{equation}\label{eq:coloronmass}
M^2_C = \frac{v^2_s}{6}(g^2_{s_1} + g^2_{s_2})~,
\end{equation}
where $g_{s_{1,2}}$ are the coupling constants of the two $SU(3)$ gauge-groups. The couplings
$g_{s_{1,2}}$ cannot be too large if the theory is to remain perturbative. Following 
\cite{Dobrescu:2009vz}, therefore, we require that the large-$N_c$ corrected loop-counting factor be less than one, 
\begin{equation}\label{eq:perturbative}
\frac{N_c\ g^2_{s_{1,2}}}{16 \pi^2}\le 1~.
\end{equation}
Using Eq. \ref{eq:coloronmass} for $N_c=3$, we then find immediately that
\begin{equation}\label{eq:massbound}
M_C \lesssim 3.0 \cdot \vevs~,
\end{equation}
and hence, from the experimental lower bound of 5.1 TeV on the coloron mass reported by CMS 
 \cite{Khachatryan:2015dcf}, we deduce that $\vevs \gtrsim 1.7$ TeV. This will have a significant impact on the model's phenomenology. For the purposes of illustration, in the rest of the paper we choose $\vevs = 2$ TeV. As we will see, one could always choose larger values of $v_s$ as well.\footnote{Values of $v_s$ smaller than 2 TeV will result, via Eq. \ref{eq:coloronmass} and the experimental lower bound of 5.1 TeV on the coloron mass, in large values of $g_{s_{1,2}}$ which can result in the scalar sector's having a Landau pole at very low energy scales. See discussion in Appendix \ref{sec:app-RGE}.}

\subsection{Fermion Sector}
\label{subsec:fermion}

As described in \cite{Chivukula:2015kua}, it is possible for the various chiralities and flavors of the standard quarks to be assigned charges under $SU(3)_{1c} \times SU(3)_{2c}$ in a range of ways, allowing for flavor-dependent and potentially chiral couplings to the colorons \cite{Frampton:1987dn,Bagger:1987fz,Hill:1991at,Chivukula:1996yr}. The model can also contain fermions beyond those identified with ordinary quarks. In particular, if the strong couplings of the ordinary fermions are taken to be chiral, additional spectator fermions will be {\it required} to cancel $SU(3)_{1c} \times SU(3)_{2c}$ anomalies.  While arbitrary generation-changing  flavor-dependent coloron couplings are strongly constrained by limits on flavor-changing neutral-currents~\cite{Chivukula:2013kw}, next-to-minimal flavor violation can be successfully implemented in a renormalizable coloron model so as to reproduce the observed fermion masses and mixings ~\cite{Chivukula:2013kw}. In what follows, therefore, we will assume that any flavor-dependent couplings are (at least to a good approximation) generation preserving, and that the subsequent coloron couplings are therefore flavor-diagonal. Furthermore, for simplicity of presentation, we will assume that both right-handed quarks of a given generation ({\it e.g.}, $t_R$ and $b_R$) have the same color properties. This last assumption can easily be relaxed in the analysis below, but unnecessarily complicates the discussion of the phenomenology at hand.

Even with the constraints described above, there are still several possibilities for assigning the color charges of the ordinary quarks. For instance, if all three generations of the ordinary quarks are chirally charged under the extended color gauge group (e.g., with all left-handed quarks charged under $SU(3)_{1c}$ and all right-handed quarks charged under $SU(3)_{2c}$), then three corresponding spectator fermion generations (carrying opposite chiral charges with respect to the quarks) are required to cancel the induced anomalies. On the other hand, if the chiral charge assignment of the third quark generation is opposite to those of the first two generations, only one additional spectator fermion generation (one up-like and one down-like spectator) is necessary. When all ordinary quarks are vectorially charged under the extended color interactions, no anomalies are induced and no spectator fermions are needed.
In the simplest cases we would generally expect there to be between zero and three chiral doublets of spectator fermions to cancel the anomalies of the extended color group.

In what follows we will consider a slight generalization of these possibilities. We will consider spectators charged as follows under $SU(3)_{1c} \times SU(3)_{2c} \times SU(2)_L \times U(1)_Y$:
\begin{itemize}
\item $N_Q$ weak doublets $Q_{L,R}$, with the $Q_L$ transforming as a $(3,1,2)_{1/6}$ and the $Q_R$ as a $(1,3,2)_{1/6}$.
\item  $n_q$ weak singlet pairs, $q_{L,R}$, with the {$q_R$} transforming as $(3,1,1)_{2/3,-1/3}$ and { $q_L$} transforming as $(1,3,1)_{2/3,-1/3}$.
\end{itemize}
With these assignments, the effective or net number of spectator doublets whose chiral charges under $SU(3)_{1c} \times SU(3)_{2c}$ help cancel the $SU(3)$ anomalies of the ordinary generations is $N_Q-n_q$. We therefore expect $0\le N_Q-n_q\le 3$. Moreover, the following Yukawa couplings give masses proportional to $\vevs$ to the spectator fermions 
\begin{equation}\label{Lferm}
- \frac{\sqrt{6}\, M_Q}{\vevs} \bar{Q}^{k}_{L} \, \Phi \, Q^{k}_{R} - \frac{\sqrt{6}\, M_q}{\vevs} \bar{q}^{\ell}_{L} \, \Phi^\dagger \, q^{\ell}_{R}
+h.c.~,
\end{equation}
where $k$ and $\ell$ index the $N_Q$ and $n_q$ families of spectators and, for convenience, we have taken each kind of spectator to be mass-degenerate.\footnote{We have also neglected additional Yukawa couplings of the form $\bar{Q}^k_L \phi q^\ell_R + h.c.$ , where $\phi$ is the Higgs field, which lead to weak-scale mixing among the various spectator fermions. Since we know that $\vevs \gg v_h$, these couplings lead to small effects which are irrelevant to the analysis given below.}

\section{The Diphoton Signal at LHC}
\label{sec:diphoton}

We will now demonstrate that the scalar $s$ or pseudoscalar ${\mathcal A}$ boson of the renormalizable coloron model could give rise to a 750 GeV diphoton resonance consistent with the signal reported from early high-energy LHC data \cite{CMS:2015dxe,ATLAS-diphoton}. Following the procedure in \cite{Chivukula:2013xka,Chivukula:2014rka}, we construct an effective Lagrangian coupling the scalar and pseudoscalar bosons to the gauge bosons (having integrated out the heavy color degrees of freedom) and the ordinary fermions. We then use this effective Lagrangian to compute the relevant production cross-sections and branching ratios. We outline the relevant computations in appendices \ref{sec:appendix-i} and \ref{sec:appendix-ii}; details may be found in \cite{Chivukula:2013xka,Chivukula:2014rka}.

In the renormalizable coloron model, the width of the $750$ GeV resonance, be it scalar or pseudoscalar, must be small.\footnote{For the scalar, we expect $\sin\chi\sim 0$ in order to be consistent with phenomenological constraints \cite{Chivukula:2013xka,Chivukula:2014rka,Chivukula:2015kua} so that both scalar and pseudscalar decays are dominated by loop induced processes and the total width must therefore be small.} Hence it is possible to evaluate the total production cross-section in the Narrow Width Approximation (NWA),
\begin{equation}
\sigma_{s,{\mathcal A}}(gg \to s,\,{\mathcal A} \to \gamma \gamma) = 
16\pi^2 \cdot {\cal N} \cdot \frac{ \Gamma_{s,\,{\mathcal A}}}{m_s} \cdot
BR(s,\,{\mathcal A} \to \gamma \gamma) \cdot BR(s,\,{\mathcal A}\to gg) \cdot \left[ \frac{d L_{gg}}{d\hat{s}}\right]_{\hat{s} = m^2_s}~.
\label{eq:simplest}
\end{equation} Here ${\cal N}$ is a ratio of spin and color counting factors which, for a color-singlet scalar produced via gluon fusion is:
\begin{equation}
{\cal N} = \frac{N_{S_s}}{N_{S_g} N_{S_g}} \cdot        
\frac{C_{s}}{C_g C_g} = \frac{1}{4}\cdot \frac{1}{64},
\end{equation}
where $N_i$ and $C_i$, respectively, count the number of spin- and color-states for the initial-state partons (denominator) and the resonance (numerator). 

Within the cross-section formula,
$L_{gg}$ is the gluon luminosity function, which we evaluate using the {\tt CTEQ6L1} parton distribution function~\cite{Pumplin:2002vw} at both 8 and 13 TeV.  In order to better match our theory predictions to the experimental results, we determine the NNLO $K$-factor using the {\tt SuSHi} program~\cite{Harlander:2012pb} in the infinite quark mass limit. We use the {\tt CT14NNLO} pdf set~\cite{Dulat:2015mca} and set  the renormalization and factorization scales to be $\mu_R=\mu_F=750 $ GeV. We find the $K$-factor to be 
 $K_{NNLO/LO}^{13 TeV}\sim 2.9$ and $K_{NNLO/LO}^{8 TeV}\sim 3.2$ and we apply this to our tree-level cross-section results to make the comparison with data more meaningful.

For both the $s$ and ${\mathcal A}$ states in the renormalizable coloron model (when $\sin\chi \approx 0$), the branching ratio to $gg$ dominates so long as all of the other scalars, colorons, and spectator fermions are heavy. In fact, $BR(s,\, {\mathcal A} \to gg) \approx 1$ so that the expression for the cross-section in Eq. \ref{eq:simplest} is proportional to $\Gamma_{s,\, {\mathcal A}} \cdot BR(s,\, {\mathcal A} \to \gamma \gamma) \approx \Gamma(s,\, {\mathcal A} \to \gamma \gamma)$. Furthermore, as shown in appendices \ref{sec:appendix-i} and \ref{sec:appendix-ii}, the partial width to diphotons is dominated by the contribution from loops of the spectator quarks $Q$ and $q$. The resonant diphoton production rate is, therefore, proportional to the square of the total number of spectator fermions $(N_Q + n_q)^2$ and inversely proportional to $\vevs^2$. Thus, as we illustrate below, for a given value of $\vevs$ some minimum number of spectators $N_Q + n_q$ will be required to make the predicted signal match the data.\footnote{The corresponding decay to $WW$ and $ZZ$ arise through a similar loop process, but in this case only
the weak-doublet spectator fermions contribute significantly -- and hence this amplitude (exactly, for $WW$, and only approximately for $ZZ$) is proportional to $N_Q$.}

\begin{figure}
	\centering
	\includegraphics[width= 0.47\textwidth]{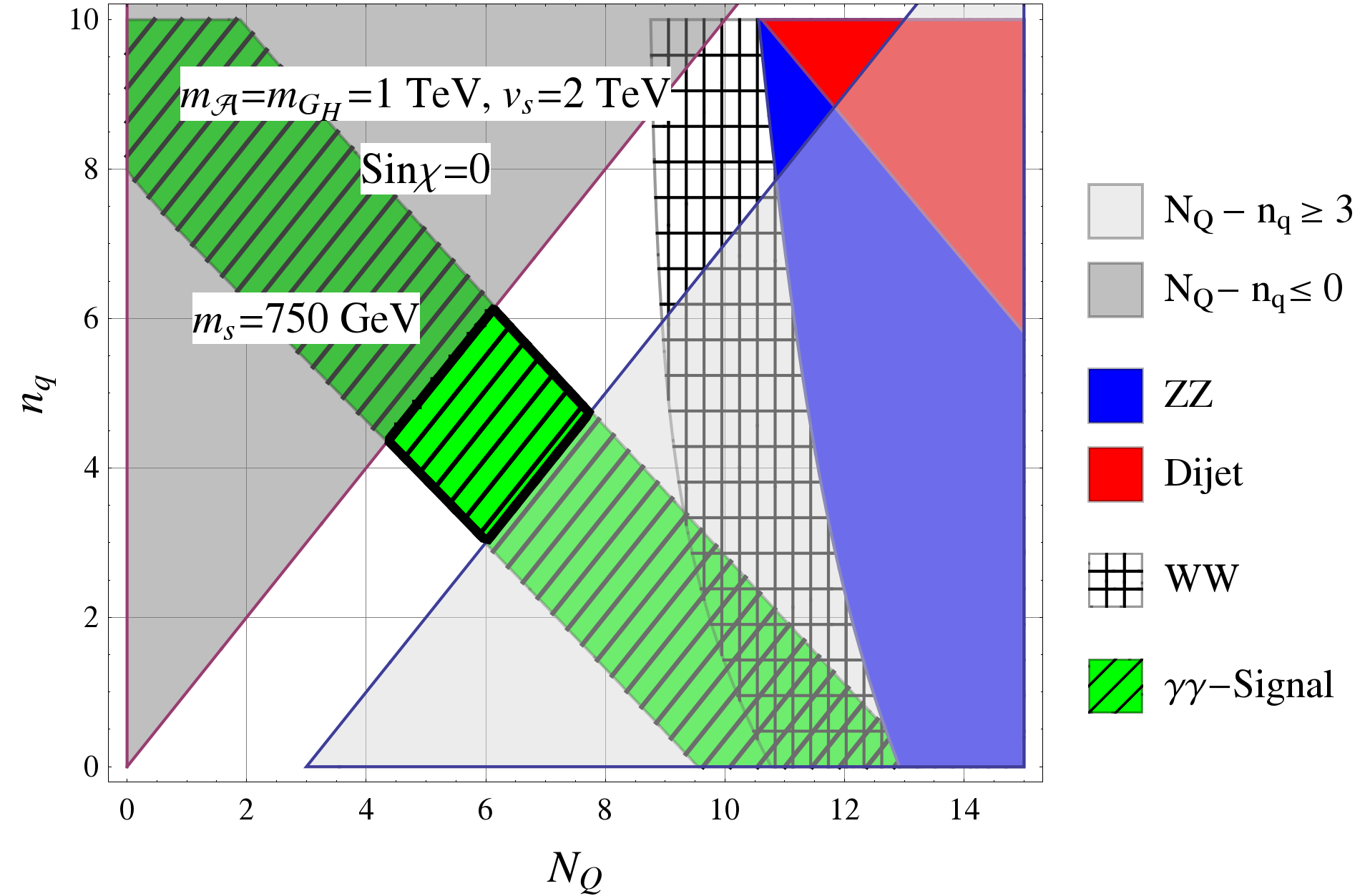}
	\includegraphics[width= 0.47\textwidth]{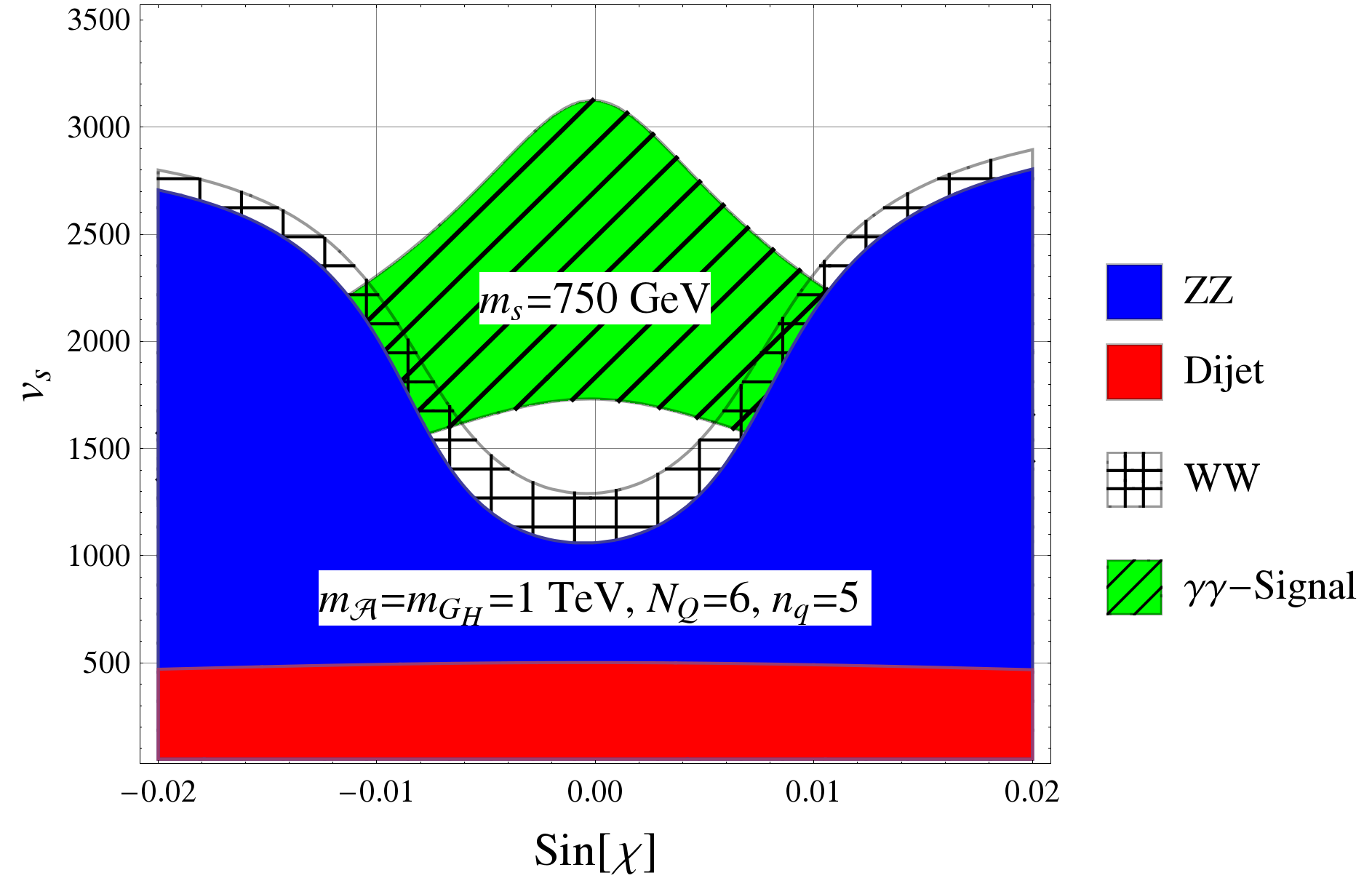}
\caption{The heavy black rim (left pane) encloses the region of the $N_Q$ vs $n_q$ plane for which the renormalizable coloron model's scalar boson $s$ is consistent with the $750$ GeV diphoton signal and other constraints.   The region shown in green (diagonally hatched) matches the 1-$\sigma$ resonance diphoton cross section of Eq. \ref{eq:signal}. Also shown are the regions excluded by $s\to WW$ searches depicted with cross-hatching \cite{Aad:2015agg}, by $s \to ZZ $ searches depicted in blue (dark gray) \cite{Aad:2015kna} and by dijet searches depicted in red (lighter gray) \cite{CMS-PAS-EXO-14-005}. The regions with translucent gray overlays correspond to values of $(N_Q,n_q)$ that are not theoretically preferred (see text for details). \textbf{Left:} Plot in the $(N_Q,n_q)$ plane for the values $\sin\chi=0,\, m_\psa =m_{G_H}=1\, {\rm TeV},\, {\rm and}\, \vevs =2\, {\rm TeV}$. Depending on the spectator fermions included, this region is sensitive to the RGE constraints discussed in Appendix~\ref{sec:app-RGE}. \textbf{Right:} Plot in the $(\sin\chi,\vevs)$ plane for 
parameter values $m_\psa =m_{G_H}=1\, {\rm TeV},\,{\rm and}\,  (N_Q, n_q) = (6,5)$. \label{fig:scalar}    }
\end{figure}

In Fig.~\ref{fig:scalar} we illustrate the region of parameter space in the renormalizable coloron model that can accomodate the observed diphoton signal  if this signal arises solely from the scalar $s$ boson. These plots are for the parameter values $ M_{Q,q} \gg 750\, {\rm GeV},\, m_\psa =m_{G_H}=1\, {\rm TeV},\, {\rm and}\, \vevs =2 \, {\rm TeV}$ -- though their appearance depends only weakly on $M_{Q,q}$, $m_\psa$ and $m_{G_H}$ so long as these particles are heavy enough to prevent $s$ from decaying to pairs of them. For $\vevs=2$ TeV and $\sin\chi=0$, the decay width to diphotons is sufficiently large to reproduce the resonance diphoton cross section of Eq. \ref{eq:signal} provided that the spectators are sufficiently numerous $(9 \lesssim N_Q + n_q \lesssim 14)$; the corresponding region is indicated in the left plot by the green (diagonally hatched) region. For larger values of $\vevs$, the required value of $N_Q + n_q$ rises proportionally. As noted in Appendix~\ref{sec:app-RGE}, the upper third of the allowed area may be excluded by the need to avoid a Landau pole in the RGE running of the weak $SU(2)$ gauge coupling.

Also plotted are the constraints arising from the non-observation of a $WW$~\cite{Aad:2015agg}, $ZZ$~\cite{Aad:2015kna}, and dijet resonance~\cite{CMS-PAS-EXO-14-005}  of the same mass. Evidently, the most difficult constraint to satisfy in this model when $\sin\chi = 0$ is simply of having a sufficiently large diphoton signal. If one increased the value of $\vevs$ and increased $N_Q + n_q$ proportionally so as to keep the signal strength in the diphoton channel constant, the minimum number of spectator quarks required to violate the dijet (Eq. \ref{eqn:sggwidth}) or diboson (e.g., Eq. \ref{eqn:sWWwidth}) bounds would also rise, leaving the model consistent with the data.

\begin{figure}
	\centering
	\includegraphics[width=0.47\textwidth]{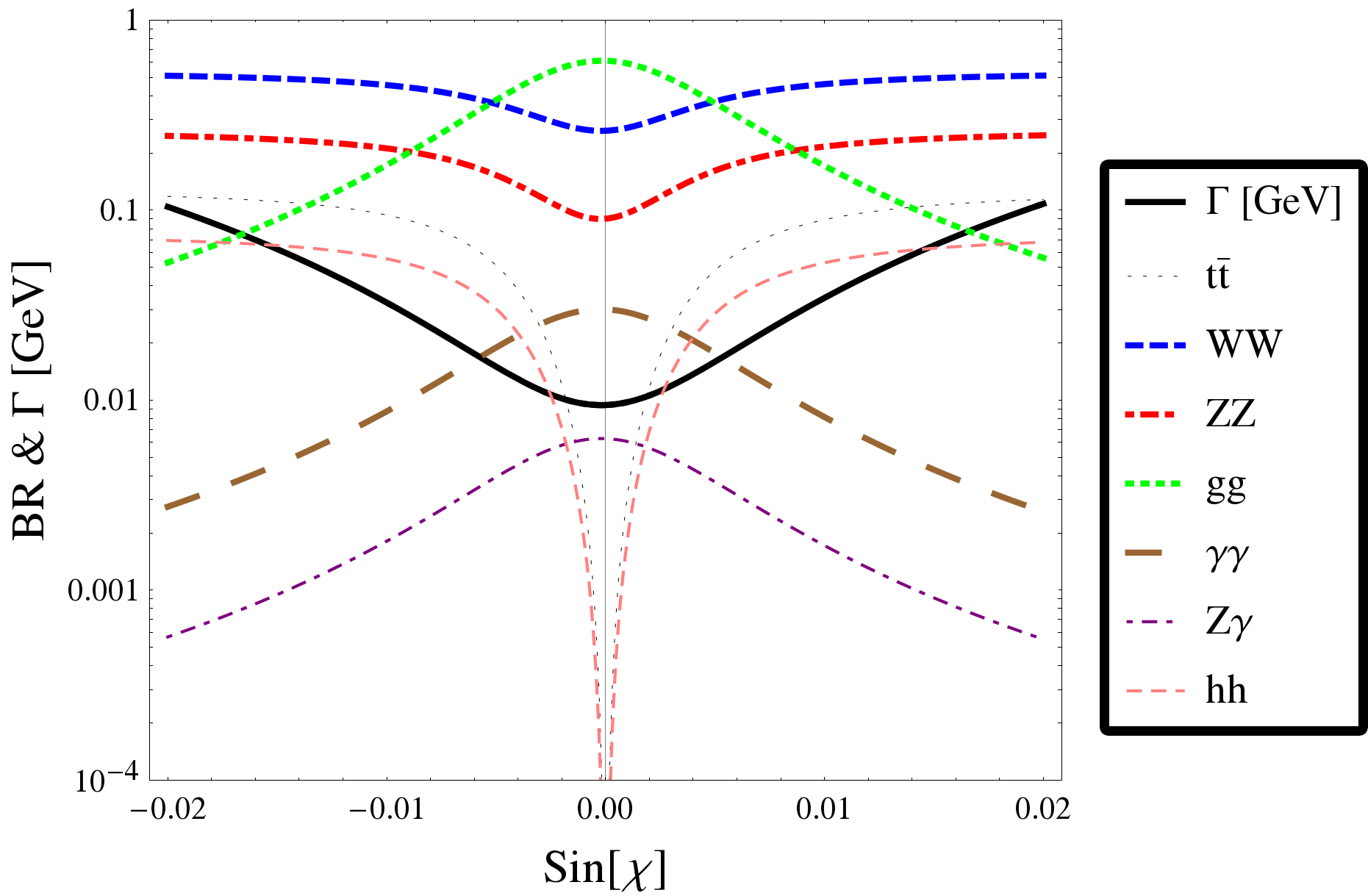}
	\caption{Scalar boson ($s$) decay width and branching ratios as a function of $\sin\chi$, for the parameter values $m_\psa = m_{G_H} = 1 \text{ TeV}$ and  $(N_Q, n_q) = (6,5)$. \label{fig:BRvariation}}
\end{figure}

If the Higgs mixing angle $\sin\chi$ is not zero, two separate effects start to suppress the branching ratio to diphotons, making it difficult to sustain a large enough signal.  First the decays to $WW$ and $ZZ$ become significant and start to cut into the available parameter space.  Second there is a destructive interference in the diphoton loop amplitude between the contributions of spectator fermions and $W$ bosons running in the loop; Fig. \ref{fig:BRvariation} illustrates that as $\sin\chi$ grows, the $WW$ and $ZZ$ widths grow while the diphoton width falls. The right hand panel of Fig. \ref{fig:scalar} demonstrates that, to be consistent with the putative signal, the $s_0 - h_0$ mixing must therefore be very small, with $|\sin\chi | \lesssim 0.01$. { Note that, as for any model in which a scalar's gaining a vacuum expectation value is the origin of the diphoton signal, a small mixing angle is not the natural consequence of any symmetry and it only occurs for a narrow range of parameters in the scalar potential.}

\begin{figure}
	\includegraphics[width= 0.45\textwidth]{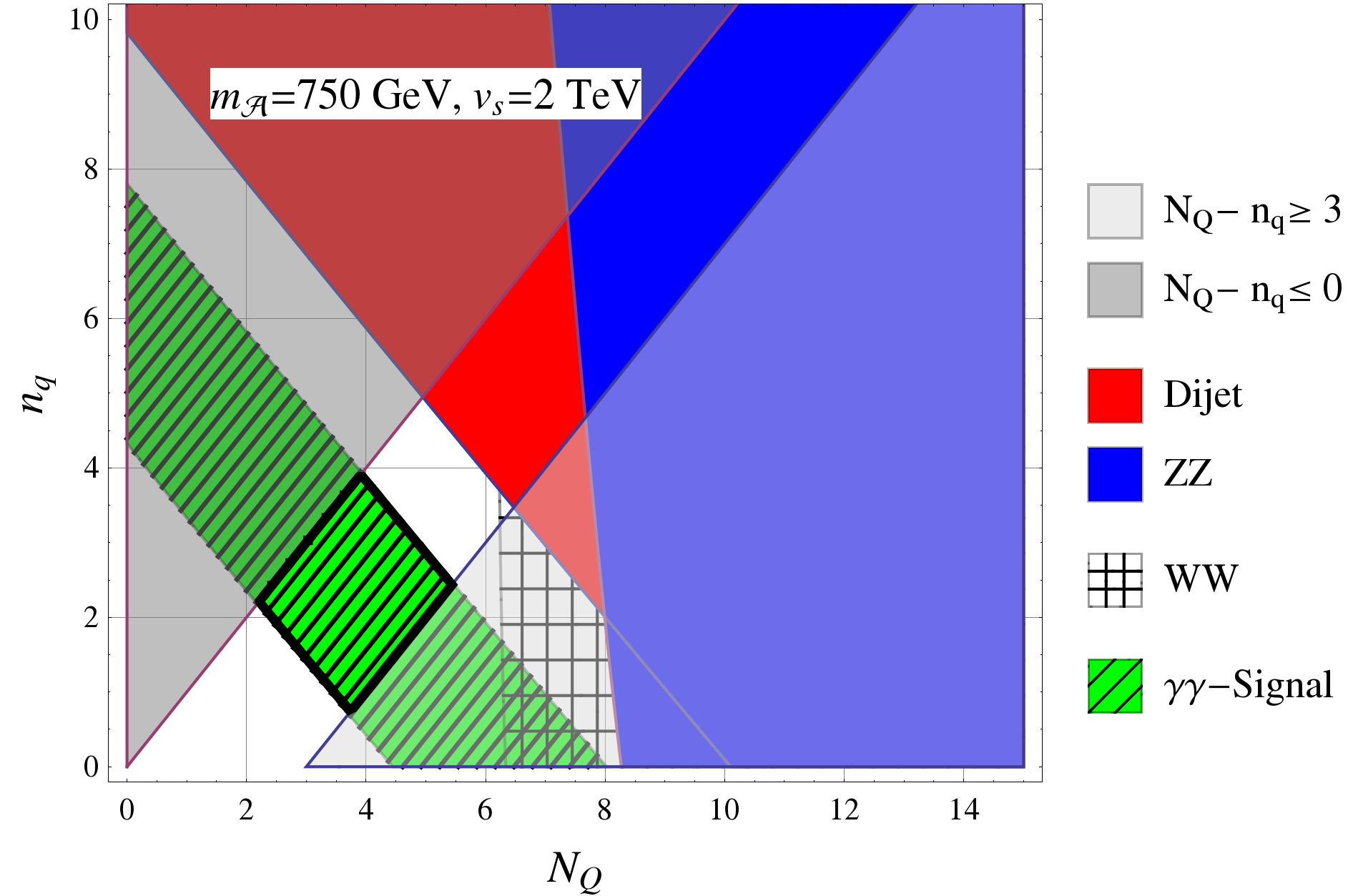} \includegraphics[width= 0.45\textwidth]{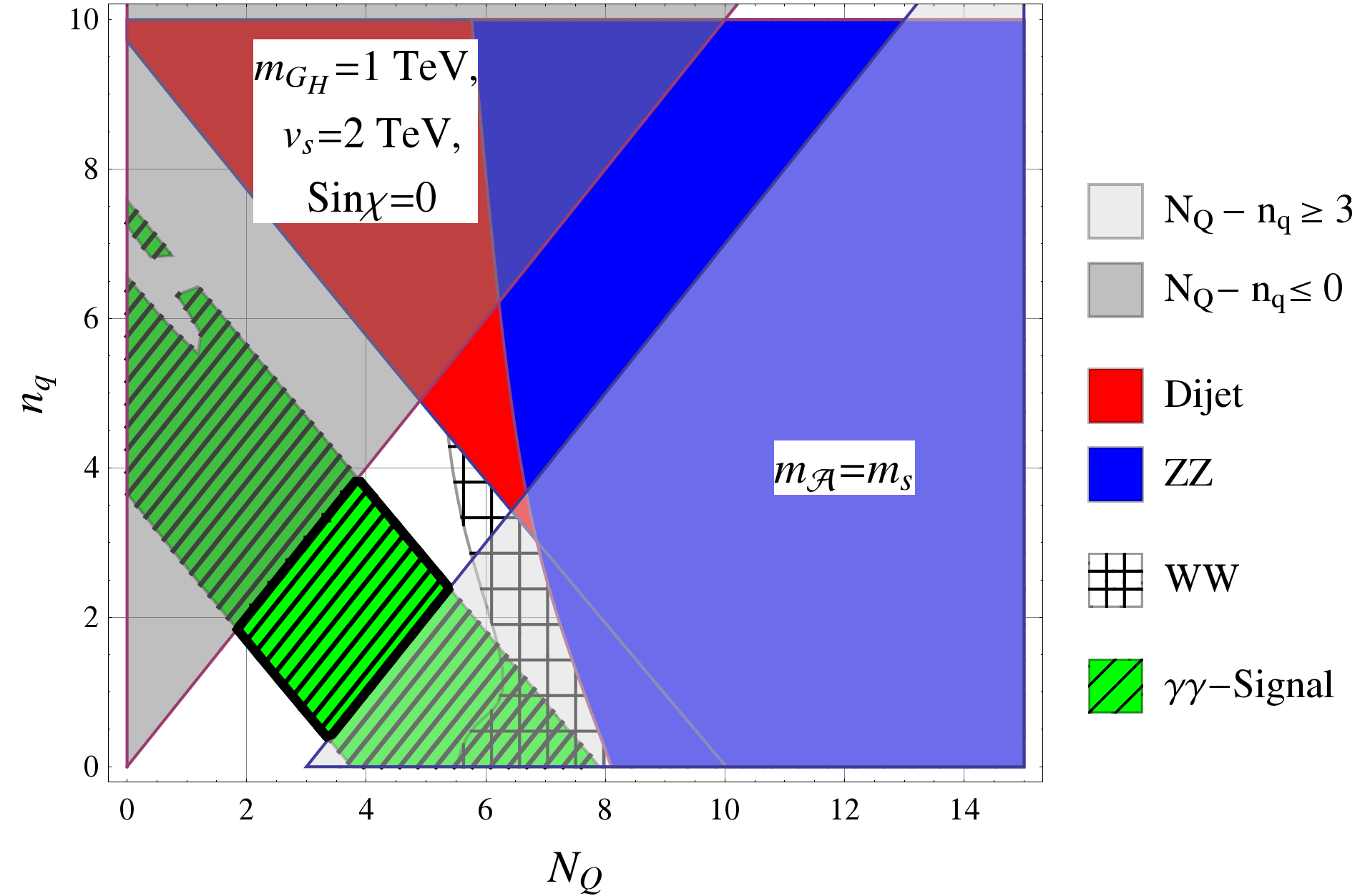}
	\caption{The heavy black rim encloses the region of the $N_Q$ vs $n_q$ plane for which the renormalizable coloron model's pseudoscalar boson ${\cal A}$ alone (left pane) or a degenerate $s,{\cal A}$ pair (right pane) is consistent with the $750$ GeV diphoton signal and other constraints. Details are as in the caption for the left pane of Fig. \ref{fig:scalar} (except that $\sin\chi$ is irrelevant for the ${\cal A}$ boson). The allowed region in each pane is unaffected by the RGE constraints in Appendix~\ref{sec:app-RGE}.
	\label{fig:pseudoscalar}}
\end{figure}

The left pane of Fig. \ref{fig:pseudoscalar} shows the region of parameter space in the renormalizable coloron model that can accomodate the dipoton signal via a 750 GeV pseudoscalar ${\mathcal A}$ boson. Here there is no $\sin\chi$ dependence, and we find that the diphoton signal can be accomodated with fewer spectator fermions, $5 \lesssim N_Q + n_q \lesssim 8$ for $\vevs=2$ TeV.\footnote{ 
This is due to the fact that the coloron and other scalars, which dominate the $s \to gg$ decay, do not contribute to $\psa \to gg$ decays. In case of the scalar, there is destructive interference between the bosonic contributions and the spectator fermion loops, so $N_Q + n_q$ is pushed toward larger values where the fermionic contribution dominates.}
As with the scalar, if $\vevs$ is increased, the total number of spectator fermions must be increased proportionally, but the constraints from non-observation of dijet and diboson decays do not become harder to satisfy. As noted in Appendix~\ref{sec:app-RGE}, the extent of the allowed region should be unaffected by the need to avoid a Landau pole in the RGE running of the gauge couplings.

Finally, in the right pane of Fig.~\ref{fig:pseudoscalar} we consider the case in which the scalar and pseudoscalar are {roughly degenerate (to within experimental resolution)},\footnote{ There is no symmetry that would enforce strict degeneracy between the scalar and psuedo-scalar resonances in this model. For other examples of models of the diphoton signal involving degenerate resonances see~\cite{Wang:2015omi,Bai:2016rmn,Djouadi:2016eyy}.} and both have masses of 750 GeV. Here we see that $4 \lesssim N_Q + n_q \lesssim 7$ can accommodate the signal when $\vevs\sim 2$ TeV.
For larger values of $\vevs$, proportionally larger values of $N_Q + n_q $ would be able to explain the diphoton signal without generating dijet or diphoton rates in excess of the bounds. Here too, as shown in Appendix~\ref{sec:app-RGE}, the extent of the allowed region should be unaffected by the need to avoid a Landau pole in the RGE running of the gauge couplings.

\begin{figure}
	\includegraphics[width= 0.45\textwidth]{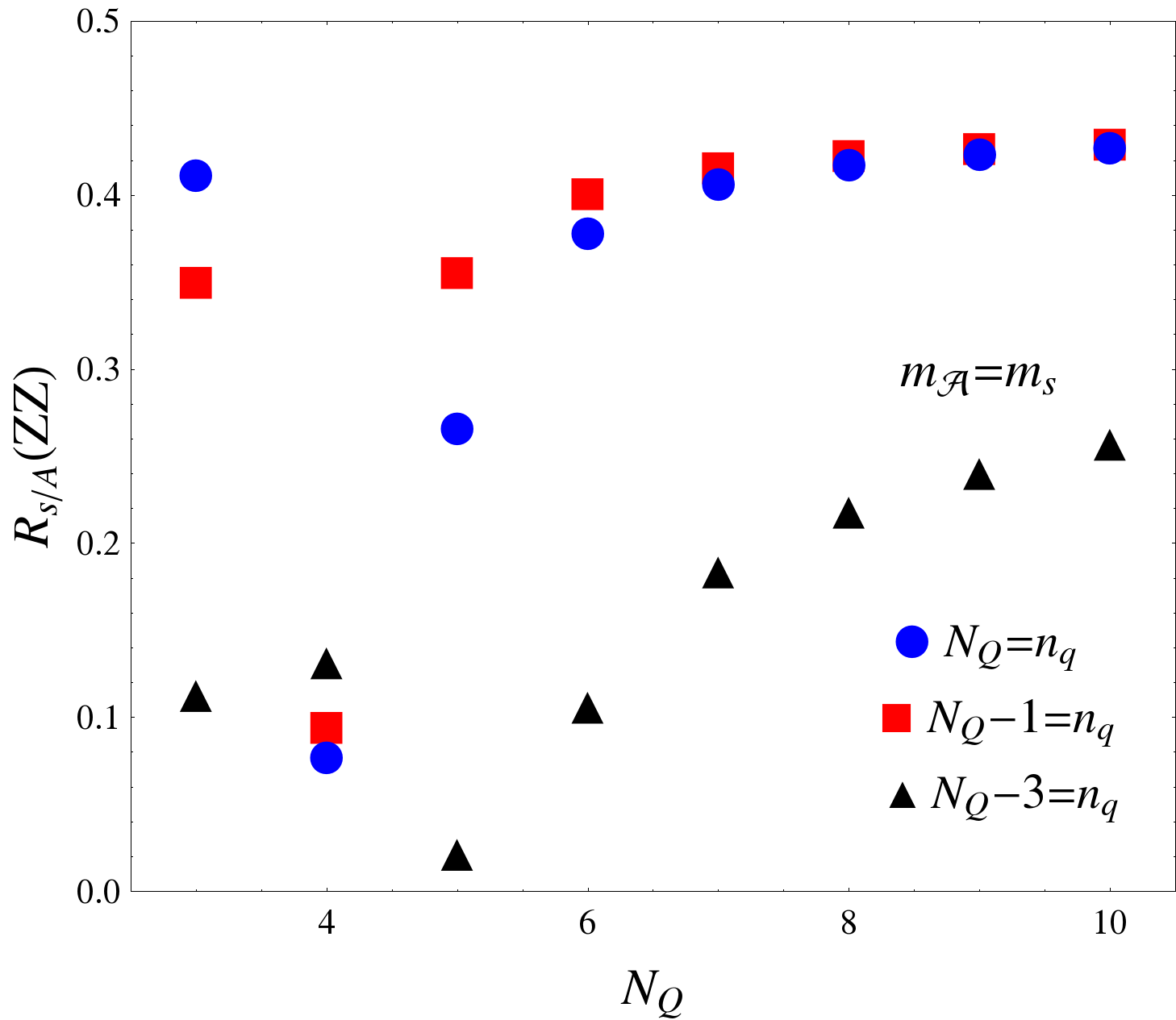} 
	\caption{Ratio of the scalar to the pseudoscalar component of the diphoton signal as a function of $N_Q$, for quasi-degenerate $m_\psa =m_s=750$ GeV. All three possible physical cases are shown $n_q=N_Q$ (blue circles), $n_q=N_Q -1$ (red squares)  and $n_q=N_Q-3$ (black triangles).\\ 
	\label{fig:pseudoscalar-ratio}
	}
\end{figure}

In the quasi-degenerate case, it is interesting to note that the pseudoscalar contribution to the diphoton rate is predicted to be larger than that of the scalar contribution to the signal. In fact, one could determine the relative sizes of the scalar and pseudoscalar components of the signal through angular observables (e.g., in $\psa ,s \to ZZ \to 4l$ decays) as a test of whether degenerate $\psa$ and $s$ were contributing (in a manner analogous to the spin-parity measurement
of the Higgs boson \cite{Khachatryan:2014kca,Aad:2015mxa}). In Fig.~\ref{fig:pseudoscalar-ratio} we show the ratio 
\begin{equation}
R_{s/\psa} = \frac{\sigma(pp \to s \to ZZ)}{\sigma(pp \to \psa \to ZZ)}\ ,
\end{equation} 
as a function of $N_Q$ for the three possible physical cases: $n_q=N_Q$, $n_q=N_Q -1$ and $n_q=N_Q-3$. 
Note that this ratio is independent of the value of $\vevs$.
The dip observed in the ratio for $N_Q\sim4,5$  is due to a cancellation between the fermion and boson (coloron and other scalar) loops that causes the $s\to gg$ branching fraction (and hence the overall scalar production cross-section) to vanish. 
Experimental determination of this ratio could help determine the value of $N_Q$ and $n_q$.\footnote{We have neglected interference effects between $s$ and $\psa$ since the decay widths of both particles are very small and since, due to the CP symmetry of the total cross-section, there is no contribution to the total cross-section from the CP-odd interference term.}

\section{Discussion}
\label{sec:conclusions}

We have shown that the scalar sector of the renormalizable coloron model can be the source of the 750 GeV resonance for which evidence has been observed at the LHC.  Either the scalar state $s$, the pseudoscalar state ${\cal A}$, or both (if degenerate) could play the role of the new diphoton resonance, while remaining consistent with precision electroweak physics and constraints from triviality and unitarity.

If the 750 GeV resonance is verified by further analysis and accumulation of more statistics, there are clear avenues for verifying that the renormalizable coloron model is the underlying new physics involved.  The most straightforward would be to look for direct evidence of the coloron resonance in dijet invariant mass or dijet angular distributions; indeed, the LHC experiments routinely look for signs of high-mass dijet resonances in each newly-collected data set  ({\it e.g.} \cite{Khachatryan:2015dcf}).  Alternatively, one could seek evidence within the LHC data for a second new spinless state ($s$ or ${\cal A}$) at a different mass or look for signs of the colored scalars $G^a_H$ as suggested in \cite{Bai:2010dj,Hill:1993hs}.  In addition, one could study other decay modes of the 750 GeV resonance to predict the expected number of spectator quarks or to differentiate between the scalar, pseudoscalar and degenerate cases discussed here. In the longer term, aspects of the model that would warrant further study would include the detailed impact of the weak-singlet spectator quarks upon the theoretical constraints on the model and the precise flavor structure of the quark sector in the presence of the various spectators.

We look forward to seeing what the next run reveals.

\begin{acknowledgments}
This material is based upon work supported by the National Science Foundation under Grant No. PHY-1519045. KM acknowledges the support of the Michigan State University High Performance Computing
Center and the Institute for Cyber Enabled Research.
\end{acknowledgments}


\appendix

\section{Couplings and Decays of the scalar}
\label{sec:appendix-i}

In this appendix and the one that follows, we outline the couplings and decay widths of the $s$ and ${\mathcal A}$ bosons in the renormalizable coloron model. We assume either the $s$ or the ${\mathcal A}$ has a mass of 750 GeV, so that it can be associated with the diphoton resonance for which evidence has been found at the LHC.  We also assume that the 750 GeV boson is the lightest new state in the spectrum, so it is kinematically forbidden to decay to non-SM particles (colorons, extra scalars, or spectator fermions). The lightest boson, then, can decay to pairs of ordinary gauge bosons or standard model quarks. We summarize the couplings of the $s$ and ${\mathcal A}$ to these SM particles in terms of an effective Lagrangian obtained by integrating out the heavy particles. For spectator quarks, we will use the heavy quark mass limit and neglect mass effects in the calculation of the decay widths of $s$ and $\psa$.

We will find that the branching ratios of the scalar particle depend sensitively on the value of $\sin\chi$,
as shown in Fig. \ref{fig:BRvariation}. As noted in the text, the diphoton branching ratio falls precipitously 
as $\sin\chi$ grows from zero -- both because the branching ratio to $WW$ and $ZZ$ grows, as well as due to destructive interference between fermion and $W$-boson loops in the diphoton decay amplitude.

\subsection{Couplings of the Scalar $s$}

The effective Lagrangian parametrizing decays of the scalar $s$ is given below:
\begin{eqnarray}
\lag^{s}_{eff}&=& 
c^{s}_{W1}\frac{2 m_W^2}{\vevh} s W_{\mu}^{+}W^{-\mu} 
+ c^{s}_{Z1}\frac{2 m_Z^2}{\vevh} s Z_{\mu}Z^{\mu} 
- c^{s}_{q_{SM}}\frac{m_{q_{SM}}}{\vevh} s q_{SM} \bar{q}_{SM}
- c^{s}_{l}\frac{m_l}{\vevh} s l \bar{l}\nn \\
&+& c^{s}_{g}\frac{\alpha_s}{12\pi \vevh} s G^{a}_{\mu\nu}G^{a\mu\nu} 
+ c^{s}_{\gamma}\frac{\alpha}{6\pi \vevh} s A^{a}_{\mu\nu}A^{a\mu\nu} 
+ c^{s}_{Z2}\frac{\alpha}{6\pi \vevh} s Z_{\mu\nu}Z^{\mu\nu} 
+ c^{s}_{W2}\frac{\alpha}{6\pi \vevh} s W^{+}_{\mu\nu}W^{-\mu\nu} \nn \\
&+& 
c^{s}_{AZ}\frac{\alpha}{6\pi \vevh} s A_{\mu\nu}Z^{\mu\nu} 
+c^{s}_{h}s h h 
+ c^{s}_{\psa}s \psa\psa 
+ c^{s}_{G_{H}}s G_{H}^{a} G_{H}^{a}\ .
\label{eq:eff-lag-s}
\end{eqnarray}
Here $W_{\mu}^{\pm}, Z_{\mu}, G^{a}_{\mu}$ and $A_{\mu}$ correspond to the $W$, $Z$, gluon and photon fields, respectively, whereas $W^{\pm\mu\nu}, Z^{\mu\nu}, G^{a\mu\nu}$ and $A^{\mu\nu}$ correspond to their field strength tensors. The $q_{SM} = \{t,b,c,s\}$ and the $l=\{\tau,\mu\}$ are the SM quark and lepton fields, respectively.\footnote{We ignore contributions from the very light first-generation fermions $u$, $d$, and $e$.}

Since the $s_0$ state does not couple to SM fermions, the rate at which the scalar mass eigenstate $s$ decays to SM quarks and leptons is determined by the mixing angle ($\sin\chi$) between the higgs ($h_0$) and scalar ($s_0$) gauge eigenstates. The same is also true for tree level decays of $s$ to a pair of massive electroweak bosons:
\begin{equation}
c^{s}_{W1}= c^{s}_{Z1}= c^{s}_{q_{SM}} = c^{s}_{l} =\sin\chi\ .
\end{equation}

On the other hand, loop-induced decays of $s$ to gauge bosons receive contributions from both non-SM and SM particles. The magnitude of each of these contributions, again, depends on the mixing between $h_0$ and $s_0$ as parametrized by $\sin\chi$.

For the coupling of gluons to $s$ we have the following expressions~\cite{Chivukula:2013xka}
\begin{eqnarray}\label{eq:cgs}
c^s_g&=&\sin\chi \,\hat{c}_g^{\text{SM}}+\cos\chi \, \hat{c}^s_g,\nonumber\\
 \nonumber\\
 \hat{c}_g^{\text{SM}}&\equiv&\, \frac{3}{4}\left(A_f(\tau^s_t)+A_f(\tau^s_b)\right)\simeq 0.4 + 1.1 i~,\nonumber\\
\hat{c}^s_g&\equiv&3\frac{v_{h}}{\vevs}\left[\frac{3}{4}A_V(\tau^s_C)- 6\left(1+\frac{m_s^2-\frac{2}{3}m_\mathcal A^2}{2m_{G_H}^2}\right)A_S(\tau^s_{G_H})+\frac{ (N_Q+ n_q)}{2}A_f(\tau^s_Q)\right],\nn\\
 &\simeq& 3\frac{\vevh}{\vevs}\left[
-\frac{21}{4} +\frac{1}{4}\left(1+\frac{m_s^2-\frac{2}{3}m_\mathcal A^2}{2m_{G_H}^2}\right) + \frac{2(N_Q + n_q)}{3}
\right]~,
\end{eqnarray}
assuming the all spectator quarks ($Q$ and $q$) are degenerate.\footnote{This need not be the case, and the formulae may be easily generlized to take this into account; in practice, however, if all spectators are much heavier than 750 GeV, all spectators contribute approximately equally.}
Here $\tau^{s}_i = \frac{m_s^2}{4 m_i^2}$. The $A_f$, $A_V$ and $A_S$ correspond to the fermionic, spin-1 and scalar loop form factors given below \cite{Chivukula:2013xka,Djouadi:2005gj}.
\begin{eqnarray}
A_{f}(\tau)&=&\frac{2}{\tau^2}(\tau + (\tau -1)f(\tau)), \nn\\
A_S(\tau)&=&\frac{\tau -f(\tau)}{8 \tau^2},\ \nn \\
A_{V}(\tau)&=&-\frac{1}{\tau^2}(2\tau^2 +3\tau +3(2\tau -1)f(\tau)),
\label{eq:formfactorae}
\end{eqnarray}
with $\tau_i\equiv\frac{m_h^2}{4m_i^2}$ and
\begin{eqnarray}
f (\tau) \equiv \left\{ \begin{array}{lc}
\mbox{arcsin}^2 \sqrt{\tau}  & \mbox{for }\tau \leq 1 \\
- \frac{1}{4} \left[ \log \frac{1+\sqrt{1-\tau^{-1}}}{1-\sqrt{1-\tau^{-1}}} - i \pi \right]^2 & 
\mbox{for }\tau > 1 \end{array} \right.\,.
\end{eqnarray}
The asymptotic limits of the loop functions
\begin{equation}
\label{eq:A-limits}
\lim_{\tau\to 0} A_f(\tau)= \frac{4}{3},\ \ 
\lim_{\tau\to 0} A_V(\tau)= -7,\ \ 
\lim_{\tau\to 0} A_S(\tau)= -\frac{1}{24}.\ \ 
\end{equation}
were used to determine the approximate expression in the last line of Eq.~\ref{eq:cgs}.

Similarly, the coupling of photons to $s$ is given by
\begin{eqnarray}\label{eq:cas}
c^s_\gamma&=&\sin\chi \,\hat{c}_\gamma^{\text{SM}}+\cos\chi \, \hat{c}^s_\gamma,\nonumber\\
 \nonumber\\
\hat{c}_\gamma^{\text{SM}}&\equiv&\,\frac{3}{4}\left[ A_V(\tau_W^s)+ \frac{4}{3}A_f(\tau^s_t)+\frac{1}{3}A_f(\tau^s_b)\right]\simeq -0.5 + 0.07i~,\nonumber\\
\hat{c}^s_\gamma&\equiv&\frac{3}{4}\frac{v_{h}}{\vevs}\left[\frac{ 5(N_Q+ n_q)}{3}A_f(\tau^s_Q)\right]\nn\\
&\simeq& \frac{\vevh}{\vevs}\frac{5(N_Q+n_q)}{3}~,
\label{eq:cg}
\end{eqnarray}
where, for simplicity, we have assumed that all spectator quarks are degenerate.
Note that, as described in the text, the contributions from gauge bosons (proportional to $\sin\chi$)
and those from spectator fermions (proportional to $\cos\chi$) interfere destructively.

The coupling of $s$ to a photon plus a $Z$ boson is
\begin{eqnarray}
c^s_{AZ}&\equiv&\sin\chi \,\hat{c}_{AZ}^{\text{SM}}+\cos\chi \,
\hat{c}^s_{AZ},\nonumber\\
 \nonumber\\
\hat{c}_{AZ}^{\text{SM}}&\simeq& 0.06 + 1.5 i~, \nn \\
\hat{c}^{s}_{AZ}&\equiv&\frac{\vevh}{\vevs}\cdot\frac{n_q(Q_u Z_u + Q_d Z_d)+ N_Q(Q_U Z_U + Q_D Z_D)}{c_W s_W}~.
\end{eqnarray}
Here $Q_i$ correspond to charges of spectators quarks, whereas $Z_i= I^{3}_{i} -Q_i s_W^2$. 

Finally, the loop-induced couplings of $s$ to $W$ and $Z$ field strengths are:
\begin{equation}
c^{s}_{W2}\equiv\frac{\vevh}{\vevs}\frac{N_Q N_c}{s_W^2}~,
\end{equation}
\begin{equation}
c^{s}_{Z2}\equiv\frac{\vevh}{\vevs}\cdot\frac{n_q(Z_u^2 + Z_d^2)+ N_Q(Z_U^2 + Z_D^2)}{c_W^2 s_W^2}~.
\end{equation}

\subsection{Decays of the Scalar $s$}

Now we will display the expressions for the decay widths of the scalar mass eigenstate $s$.

\subsubsection{Decays to fermions: $s \to f \bar{f}$ }

The decay width to SM fermions is given by
\begin{equation}
\Gamma(s\to f\bar{f}) =\sin^2\chi\frac{ N_c}{8\vevh^2\pi} m_{s} m_{f}^2 \beta_{f}^3~.
\end{equation}
Here $\beta_f = (1 - 4m_f^2/m_{s}^2)^{1/2}$.

\subsubsection{ Decay to a pair of photons: $s \to \gamma \gamma$}

The decay width to photons proceeds via loops of spectator quarks, SM quarks and $W$ bosons, yielding
\begin{gather}
\Gamma(s \to \gamma \gamma)= 
\frac{\alpha ^2 m_s^3} {256 \pi ^3 \vevh^2}
\left| 
 \cos\chi (n_q+N_Q)\frac{\vevh}{\vevs}N_c\cdot\frac{5}{9}\cdot\frac{4}{3}+
\sin\chi\sum_{f=b,t}N_c Q_f^2 A_f(\tau_{f}^{s})
 +\sin\chi A_V(\tau_{W}^{s})
\right| ^2.
\end{gather}
using the form factors defined above in Eq.~\ref{eq:formfactorae} and the $\tau \to 0$ limit of $A_f(\tau)$ for the heavy spectator fermion contribution.
Note that for $\sin\chi=0$, the diphoton amplitude is proportional to $(N_Q+n_q)/\vevs$,
as discussed in the text.

\subsubsection{ Decays to a pair of gluons: $s \to g g$}
Loop-induced decays of $s$ to a pair of gluons proceed through spectator quarks, colorons, color octet scalars, and SM quarks,
yielding,
\begin{eqnarray}
\Gamma(s \to g g)&=& 
\frac{\alpha_s ^2 m_s^3} {72 \pi ^3 \vevh^2}
\bigg| 
\sin\chi\sum_{f=b,t} \frac{3}{4} A_f(\tau_{f}^{s})
+2 \cos\chi \frac{\vevh}{\vevs} (n_q+N_Q) \nn \\
&&+\frac{9}{4}\cos\chi\frac{\vevh}{\vevs}A_V(\tau^s_{C})
-18\cos\chi\frac{\vevh}{\vevs}
\left(
1 + \frac{m_s^2 -\frac{2}{3}m_\psa^2}{2 m_{G_H}^2}
\right)A_S(\tau^s_{G_H})
\bigg| ^2 \ .  \label{eqn:sggwidth}
\end{eqnarray}
Since the form factors converge quickly, we use the heavy mass limit given in  Eq.~\ref{eq:A-limits}.

\subsubsection{ Decay to a photon plus a Z-boson: $s \to Z \gamma$}

Similiarly, loop-induced decays yield the $Z\gamma$ partial-width
\begin{eqnarray}
\Gamma(s \to Z \gamma)&=& \frac{G_{\mu}^2 m_W^2\alpha m_{s}^3}{64 \pi^4}\left(1 - \frac{m_Z^2}{m_s^2}\right)^3  \times
\bigg|
\sum_{f = b,t} \sin\chi Q_f N_c \frac{ 2I_3^f -4s_W^2 Q_f}{c_W}A_{f}(\bar{\tau}_f,\bar{\lambda}_f)\nn \\ &+&  \sin\chi A_{W}(\bar{\tau}_W,\bar{\lambda}_W)\nn \\
&+& \cos \chi \frac{2}{9c_W} \left(
 N_Q(9 - 10 s_W^2) - 10 n_q s_W^2 \right)
\bigg|^2~.
\end{eqnarray}
The loop functions appearing in this expression are \cite{Djouadi:2005gj}.
\begin{eqnarray}
A_f(\bar{\tau}_f,\bar{\lambda}_f) &=&   I_1(\bar{\tau}_f,\bar{\lambda}_f) - I_2(\bar{\tau},\bar{\lambda})~,\nonumber \\
A_W(\bar{\tau}_W,\bar{\lambda}_W) &=& c_W \left[ \left(1+\frac{2}{\bar{\tau}_W}\right)\frac{s_W^2}{c_W^2}   -  \left(5+\frac{2}{\bar{\tau}_W}\right) 
\right] I_1(\bar{\tau}_W,\bar{\lambda}_W) \nonumber \\
&+&  4 c_W \left( 3 - \frac{s_W^2}{c_W^2} \right) I_2(\bar{\tau}_W,\bar{\lambda}_W)~,
\end{eqnarray}
with
\begin{eqnarray}
I_1(\bar{\tau},\bar{\lambda}) &\equiv& \frac{\bar{\tau} \bar{\lambda}}{2(\bar{\tau}-\bar{\lambda})}
+\frac{\bar{\tau}^2 \bar{\lambda}}{2(\bar{\tau}-\bar{\lambda})^2}
\Big( \bar{\lambda} \left[f(\bar{\tau})-f(\bar{\lambda})\right] 
+ 2 \left[g(\bar{\tau})-g(\bar{\lambda})\right] \Big)~,
\nonumber \\
I_2(\bar{\tau},\bar{\lambda}) &\equiv&  -\frac{\bar{\tau} \bar{\lambda}}{2(\bar{\tau}-\bar{\lambda})}
\left[f(\bar{\tau})-f(\bar{\lambda})\right] ~.
\end{eqnarray}
Note that $\bar{\lambda}\equiv 4m^2/m_Z^2$ and $\bar{\tau} \equiv 4m^2/m_{h}^2$, 
and the functions $f$ and $g$ are defined by 
\begin{equation}
f(\bar{\tau}) \equiv \left\{ \begin{array}{ll}
{\rm arcsin}^2 \sqrt{1/\bar{\tau}} & \bar{\tau} \geq 1 \\
-\frac{1}{4} \left[ \log \frac{1 + \sqrt{1-\bar{\tau} } }
{1 - \sqrt{1-\bar{\tau}} } - i \pi \right]^2 \ \ \ & \bar{\tau} <1
\end{array} \right. ,
\end{equation}
\begin{equation}
g(\bar{\tau}) \equiv \left\{ \begin{array}{ll}
\sqrt{\bar{\tau}-1} \ {\rm arcsin} \sqrt{1/\bar{\tau}} & \bar{\tau} \geq 1 \\
\frac{1}{2} \sqrt{1-\bar{\tau}} \left[ \log \frac{1 + \sqrt{1-\bar{\tau} } }
{1 - \sqrt{1-\bar{\tau}} } - i \pi \right]~ \ \ \ & \bar{\tau} <1
\end{array} \right. .
\end{equation}
For the term involving spectator quarks (and proportional to $\cos\chi$) we have used the asymptotic limit 
\begin{equation}
\lim_{\bar{\tau},\bar{\lambda} \to \infty} A_f(\bar{\tau},\bar{\lambda})= \frac{1}{3}.
\end{equation}

\subsubsection{ Decay to a pair of massive gauge bosons: $s \to VV$}

The decays of $s$ to pairs of massive electroweak bosons can proceed both at tree-level, largely due to
longitudinally-polarized particles (for $\sin\chi \neq 0$), or through loop-level processes mainly due to
transversely-polarized particles. The total decay width to $ZZ$ is 
\begin{eqnarray}
\Gamma(s &\to& ZZ) = \sin^2 \chi\frac{m_s^3}{32 \vevh^2 \pi}\sqrt{1 - 4 x}(1 - 4 x + 12 x^2) \nn\\
&+& \cos^2\chi \frac{m_s^3  \alpha^2} {144 c_W^4  s_W^4  \pi^3 \vevs^2}N_c^2 (N_Q (Z_D^2 + Z_U^2) + 
n_q (Z_d^2 + Z_u^2))^2
\sqrt{1 - 4 x}(1 - 4 x + 6 x^2)\nn \\
&+&\sin\chi \cos\chi \frac{\vevh}{\vevs} \frac{N_c m_s \alpha^2 }{8 \pi c_W^4 s_W^4} (N_Q (Z_D^2 + Z_U^2) + 
n_q (Z_d^2 + Z_u^2)) \sqrt{1- 4x}(1 -2x)\ ,
\end{eqnarray}
where $x=m_Z^2/m_s^2$ and $Z_i = I^3_i - Q_i s_W^2$. 

Similarly, the partial width to $WW$
is given by
\begin{eqnarray}
\Gamma(s \to W^{+}W^{-}) &=& \sin^2 \chi\frac{2 m_s^3}{32 \vevh^2 \pi}\sqrt{1 - 4 x}(1 - 4 x + 12 x^2) \nn\\
&+& \cos^2\chi \frac{m_s^3  \alpha^2} {288  s_W^4  \pi^3 \vevs^2}N_c^2 N_Q^2 \sqrt{1 - 4 x}(1 - 4 x + 6 x^2)\nn \\
&+&\sin\chi \cos\chi \frac{\vevh}{\vevs} \frac{N_c m_s N_Q \alpha^2 }{8 \pi  s_W^4}\sqrt{1- 4x}(1 -2x) \ , \label{eqn:sWWwidth}
\end{eqnarray}
where $x=m_W^2/m_s^2$.


\subsubsection{Decays $s \to h h$} 

Finally, the scalar coupling coefficients derived in~\cite{Chivukula:2013xka}, yield
\begin{eqnarray}
c_h^s
&=&-\frac{\sin\chi\cos\chi}{2v_h\vevs}\left[v_h\left(\frac{m_{\mathcal{A}}^2}{3}+2m_h^2+m_s^2\right)\sin\chi+\vevs(2m_h^2+m_s^2)\cos\chi\right]~,\nonumber\\
c_\mathcal{A}^s
&=&-\frac{m_\mathcal{A}^2+m_s^2}{2\vevs}\cos\chi~,\nonumber\\
c_{G_H}^s&=&-\frac{m_s^2+2m_{G_H}^2-\frac{2}{3}m_\mathcal{A}^2}{2\vevs}\cos\chi~.
\end{eqnarray}
The associated decay widths for $i=h, \mathcal{A}, G_H^a$ are
\begin{equation}
\Gamma(s\to i i)=\frac{(c_i^s)^2}{8\pi m_s}\sqrt{1-\frac{4m_i^2}{m_s^2}},\quad
\end{equation}
with $c_i^s=c_h^s, c_\mathcal{A}^s, c_{G_H}^s$ respectively. For the range of parameters examined here, only the $hh$ amplitude is relevant to our analyses.

\section{Couplings and Decays of pseudoscalar $\psa$}
\label{sec:appendix-ii}

The effect of integrating out a heavy fermion loop upon pseudoscalar decays can be estimated by adding a contribution related to the ABJ anomaly~\cite{Adler:1969gk,Bell:1969ts}.
The effective Lagrangian parametrizing decays of the pseudoscalar is shown below.

\begin{eqnarray}
\lag^{\psa}_{eff}&=& 
\frac{(N_Q + n_q)\alpha_s}{4\pi \vevs} \psa G^{a}_{\mu\nu}G^{a\mu\nu} 
+ \left[n_q(Q_u^2 + Q_d^2)+ N_Q(Q_U^2 + Q_D^2)\right]\frac{N_c\alpha}{4\pi \vevs} \psa A^{a}_{\mu\nu}\tilde{A}^{a\mu\nu} \nn \\
&+& \left[n_q(Z_u^2 + Z_d^2)+ N_Q(Z_U^2 + Z_D^2)\right] \frac{N_c\alpha}{4\pi s_W^2 c_W^2 \vevs} \psa Z_{\mu\nu}\tilde{Z}^{\mu\nu} \nn \\
&+& \left[n_q(Q_u Z_u + Q_d Z_d)+ N_Q(Q_U Z_U + Q_D Z_D)\right] \frac{N_c\alpha}{4\pi s_W c_W \vevs} \psa Z_{\mu\nu}\tilde{A}^{\mu\nu} \nn \\
&+&\frac{N_Q N_c\alpha}{4\pi s_W^2 \vevs} \psa W^{+}_{\mu\nu}\tilde{W}^{-\mu\nu}~. \nn \\
\label{eq:eff-lag-A}
\end{eqnarray}
Here the $Q_i$ correspond to electric charges of spectators quarks, while the $Z_i= I^{3}_{i} -Q_i s_W^2$ are the couplings to the $Z$-boson. 

Using the effective Lagrangian, we then find the decay widths
\begin{equation}
\Gamma(\psa \to \gamma \gamma) = \left[ n_q(Q_u^2 + Q_d^2) + 
N_Q (Q_D^2 + Q_U^2) \right]^2\frac{ m_\psa^3 N_c^2 \alpha^2}{64 \pi^3 \vevs^2}~,
\end{equation}
\begin{equation}
\Gamma(\psa \to gg) = \left[2(n_q + N_Q) \right]^2\frac{m_\psa^3  \alpha_s^2}{32 \pi^3 \vevs^2}~,
\end{equation}
\begin{equation}
\Gamma(\psa \to Z \gamma) = \left[ n_q(Q_u Z_u + Q_d Z_d) + 
N_Q (Q_D Z_D + Q_U Z_U) \right]^2 
\left(
1 - \frac{m_Z^2}{m_\psa^2}
\right)^{\frac{3}{2}}
\frac{ m_\psa^3 N_c^2 \alpha^2}{128 s_W^2 c_W^2 \pi^3 \vevs^2}~,
\end{equation}
\begin{equation}
\Gamma(\psa \to Z Z) = \left[ n_q(Z_u^2 + Z_d^2) + 
N_Q (Z_D^2 + Z_U^2) \right]^2
\left(
1 - 4\frac{m_Z^2}{m_\psa^2}
\right)^{\frac{3}{2}}
\frac{ m_\psa^3 N_c^2 \alpha^2}{64 s_W^4 c_W^4 \pi^3 \vevs^2}~,
\end{equation}
\begin{equation}
\Gamma(\psa \to W W) = 
2\frac{N_Q^2 m_\psa^3 N_c^2 \alpha^2}{256 s_W^4 \pi^3 \vevs^2}
\left(
1 - \frac{m_W^2}{m_\psa^2}
\right)^{\frac{3}{2}}~.
\end{equation}

\section{Renormalization Group Evolution of Couplings}
\label{sec:app-RGE}
In this appendix, we write down the RGE equations of the gauge couplings of the model and discuss their evolution. 
The renormalization group equations are given by the following expressions:
\begin{equation}\label{bgauge}
\begin{split}
\pbrac{4\pi}^{2}\beta_{g} =& -g^{3} \tbrac{+\frac{19}{6}  - 2 N_{Q}}\ , \\
\pbrac{4\pi}^{2}\beta_{g'} =& +g'^{\,3} \tbrac{+\frac{41}{6} +\frac{2}{9} N_{Q} +\frac{20}{9} n_{q}} \ , \\
\pbrac{4\pi}^{2}\beta_{g_{s_{1}}} =& -g_{s_{1}}^{3} \tbrac{\begin{cases}
	9&(\text{for}\, N_{q}\neq 0)\\
	7&(\text{for}\, N_{q}=0)
	\end{cases} \, - \frac{2}{3}N_{Q}  - \frac{2}{3}n_{q} -\frac{1}{2} }\ , \quad \\
\pbrac{4\pi}^{2}\beta_{g_{s_{2}}} =&-g_{s_{2}}^{3} \tbrac{\begin{cases}
	9&(\text{for}\, N_{q}\neq 0)\\
	11&(\text{for}\, N_{q}=0)
	\end{cases} \, - \frac{2}{3} N_{Q} - \frac{2}{3} n_{q} -\frac{1}{2} } \ .
\end{split}
\end{equation}

\begin{equation}\label{bYukawa}
\begin{split}
\pbrac{4\pi}^{2}\beta_{y_{t}} =&\, y_{t} \tbrac{ \begin{cases}
	-4\pbrac{g^{2}_{s_{1}}+g^{2}_{s_{2}}}&(\text{for}\, N_{q}\neq 0)\\
	-8 g^{2}_{s_{1}}&(\text{for}\, N_{q}=0)
	\end{cases} \, - \frac{9}{4} g^{2} - \frac{17}{12} g'^{\,2} +\frac{9}{2} y_{t}^{2} }\ , \\
\pbrac{4\pi}^{2}\beta_{Y_{Q}} =&\, Y_{Q} \tbrac{ - 4 \pbrac{g^{2}_{s_{1}} + g^{2}_{s_{2}}} - \frac{9}{2} g^{2} -\frac{1}{6} g'^{\, 2} + \pbrac{3+2N_{Q}} Y_{Q}^{2} + n_{q} (y_{q_u}^2 + y_{q_d}^2 )} \ , \\
\pbrac{4\pi}^{2}\beta_{y_{q_u}} =&\, y_{q_u} \tbrac{ - 4 \pbrac{g^{2}_{s_{1}} + g^{2}_{s_{2}}}  -\frac{8}{3} g'^{\, 2} + \pbrac{3+n_{q}} y_{q_u}^{2}  + n_{q} y_{q_d}^{2} + 2 N_{Q} Y_{Q}^2}\ , \\
\pbrac{4\pi}^{2}\beta_{y_{q_d}} =&\, y_{q_d} \tbrac{ - 4 \pbrac{g^{2}_{s_{1}} + g^{2}_{s_{2}}}  -\frac{2}{3} g'^{\, 2} + \pbrac{3+n_{q}} y_{q_d}^{2}  + n_{q} y_{q_u}^{2} + 2 N_{Q} Y_{Q}^2}\ .
\end{split}
\end{equation}

\begin{equation}\label{bquartic}
\begin{split}
\pbrac{4\pi}^{2}\beta_{\lambda_{h}} =&\, +4\lambda_{h}^{2} + 54 \lambda_{m}^{2} +3\lambda_{h} \tbrac{4 y_{t}^{2} - 3g^{2} - g'^{2}}- \frac{9}{4} \tbrac{16 y_{t}^{4} - 2 g^{4} - (g^{2} + g'^{2})^{2}}  \ , \\
\pbrac{4\pi}^{2}\beta_{\lambda_{m}} =&\, \lambda_{m} \bigg[ +4\lambda_{m} + 2\lambda_{h} +\frac{20}{3}\lambda_{s}' + \frac{16}{3} \kappa_{s} + \frac{3}{2} \tbrac{4 y_{t}^{2} - 3g^{2} - g'^{2}} \\ &+ 4\tbrac{ N_{Q} \,Y_{Q}^{2} +\, \frac{n_{q}}{2}( y_{q_u}^{2} + y_{q_d}^{2}) - 2 \pbrac{g^{2}_{s_{1}} + g^{2}_{s_{2}}}}\bigg] \ , \\
\pbrac{4\pi}^{2}\beta_{\lambda_{s}'} =&\, +\frac{26}{3}\lambda_{s}'^{\, 2} +12\lambda_{m}^{2} +\frac{32}{3}\kappa_{s}^{2} +\frac{32}{3}\lambda_{s}' \kappa_{s} + 8\lambda_{s}' \tbrac{ N_{Q} \,Y_{Q}^{2} + \frac{n_{q}}{2}( y_{q_u}^{2} + y_{q_d}^{2})  - 2 \pbrac{g^{2}_{s_{1}} + g^{2}_{s_{2}}}} \\ &- 8\tbrac{ N_{Q} \, Y_{Q}^{4} + \frac{n_{q}}{2}( y_{q_u}^{4} + y_{q_d}^{4})  -\pbrac{g^{2}_{s_{1}} + g^{2}_{s_{2}}}^{2}} \ , \\
\pbrac{4\pi}^{2} \beta_{\kappa_{s}} =&\,  +8\kappa_{s}^{2} + 4 \kappa_{s} \lambda_{s}' + 8\kappa_{s} \tbrac{ N_{Q} \,Y_{Q}^{2}  +\frac{n_{q}}{2}( y_{q_u}^{2} + y_{q_d}^{2}) - 2 \pbrac{g^{2}_{s_{1}} + g^{2}_{s_{2}}}} \\ &- 4\tbrac{ 2N_{Q} \, Y_{Q}^{4} +  n_{q} \, (y_{q_u}^{4} + y_{q_d}^{4}) -\frac{5}{8}\pbrac{g^{4}_{s_{1}} + g^{4}_{s_{2}}} + g^{2}_{s_{1}} g^{2}_{s_{2}}} \ .
\end{split}
\end{equation}
%
In order for our model to explain the observed diphoton signal, both $n_q$ and $N_Q$ must be greater than zero.
We therefore only consider the running of the couplings for the cases where $N_{Q} \ne 0$ and $n_Q \ne 0$. Furthermore, we recall that there are three possibilities depending on the number of chiral generations, $N_q\equiv N_Q - n_q=\{0,1,3\}$. We assume the SM quark charge assignments are as given in Ref.~\cite{Chivukula:2015kua}, namely either all vectorially charged under $SU(3)_{1c}$ ($N_q=0$), or all three generations chirally charged  under $SU(3)_{1c}\times SU(3)_{2c}$ ($N_q \neq 0$).

Since this is a theory of fundamental scalar particles, we expect that the couplings (in particular the scalar self-couplings) will have a Landau pole at high energies. The theory must therefore be considered a low-energy effective theory valid only below the energy scale of the Landau pole, above which a more fundamental high-energy theory must be found. In order for our phenomenological investigation of the low-energy theory to be valid, we require that the scalar low-energy theory not have a Landau pole at energies below 10 TeV. This requirement results in constraints on the parameters -- in particular, on the value of the scalar vev $v_s$. 

The strongest constraints on the validity of the effective theory arise from the running of the scalar self coupling $\lambda_{s}'$. The most dangerous term in $\beta_{\lambda'_s}$ comes from the coloron gauge contributions, and is proportional to $(g^2_{s_1}+g^2_{s_2})^2$. 
As noted below Eq. \ref{eq:perturbative}, the experimental coloron mass bound of 5.1 TeV results in a value of $v_s >1.7$ TeV -- however a value of $v_s$ this low results in a Landau Pole for $\lambda'_s$ below 10 TeV.
For our choice of $\vevs = 2$ TeV, there are no Landau poles up to a scale of $10$~TeV, for the entire region of parameter space for the pseudoscalar and degenerate case (see Fig.~\ref{fig:pseudoscalar}). However for the scalar case, we find that we require $N_Q + n_q < 10 $ in order to avoid Landau poles below a scale of $10$~TeV. 

For any fixed coloron mass, this situation can be quite easily mitigated. Choosing a larger $\vevs$  (and larger values of $N_Q + n_q$ to accomodate the diphoton signal) can easily push the Landau pole, and hence the validity of this effective theory, to much higher scales and all three cases-- scalar, pseudoscalar and degenerate -- can still explain the diphoton excess.

\bibliography{diphoton}
\bibliographystyle{apsrev}

\end{document}